\documentclass[12pt,showpacs,preprintnumbers,amsmath,amssymb]{revtex4}
\usepackage{graphicx}
\usepackage{epsfig}
\usepackage{dcolumn}
\usepackage{bm}
\begin{document}        

\title{\bf Study of the process $\bf e^+e^- \to \mu^+\mu^-$ in the energy
           region  $\bf \sqrt{s}=980$, 1040 -- 1380 MeV.}
\author{ M.N.Achasov}\email{achasov@inp.nsk.su}
\author{ V.M.Aulchenko}
\author{ K.I.Beloborodov}
\author{A.V.Berdyugin}
\author{A.G.Bogdanchikov}
\author{\fbox{A.D.Bukin}}
\author{D.A.Bukin}
\author{T.V.Dimova}
\author{V.P.Druzhinin}
\author{V.B.Golubev}
\author{I.A.Koop}
\author{A.A.Korol}
\author{S.V.Koshuba}
\author{A.P.Lysenko}
\author{E.V.Pakhtusova}
\author{E.A.Perevedentsev}
\author{S.I.Serednyakov}
\author{Yu.M.Shatunov}
\author{Z.K.Silagadze}
\author{A.N.Skrinsky}
\author{Yu.A.Tikhonov}
\author{A.V.Vasiljev}
\affiliation{
         Budker Institute of Nuclear Physics,  \\
         Siberian Branch of the Russian Academy of Sciences \\
         11 Lavrentyev,Novosibirsk,630090, Russia \\
         Novosibirsk State University, \\
         630090, Novosibirsk, Russia}					     

\begin{abstract}
 The cross section of the process $e^+e^-\to\mu^+\mu^-$ was measured in the
 SND experiment at the VEPP-2M $e^+e^-$ collider in the energy region
 $\sqrt{s}=980,$ 1040 -- 1380 MeV. The event numbers of the process 
 $e^+e^-\to\mu^+\mu^-$ were normalized to the integrated luminosity measured
 using $e^+e^-\to e^+e^-$ and $e^+e^-\to\gamma\gamma$ processes. The ratio 
 of the measured cross section to the theoretically predicted value is 
 $1.006\pm 0.007 \pm 0.016$ and $1.005 \pm 0.007 \pm 0.018$ in the first and
 second case respectively. Using results of the measurements, the
 electromagnetic running coupling constant $\alpha$ in the energy region
 $\sqrt{s}=1040$ -- 1380 MeV was obtained $<1/\alpha> = 134.1\pm 0.5 \pm 1.2$
 and this is in agreement with theoretical expectation.
\end{abstract}


\maketitle

\section{Introduction}

 The process $e^+e^-\to\mu^+\mu^-$ is the simplest process in the electroweak
 theory and at the same time it constitutes an important tool in the high 
 energy physics. It plays a fundamental role for studies of the 
 electromagnetic and weak interactions, electromagnetic properties of hadrons.
 This process was used for quantum electrodynamics (QED) tests, in electroweak
 interference studies, in leptonic width  measurements of the
 $I^{G}J^{PC}=0^{-}1^{--}$ vector mesons and $Z$-bozon, for the study of the 
 running electromagnetic coupling constant $\alpha(s)$.
 
\begin{figure}[t]
\epsfig{figure=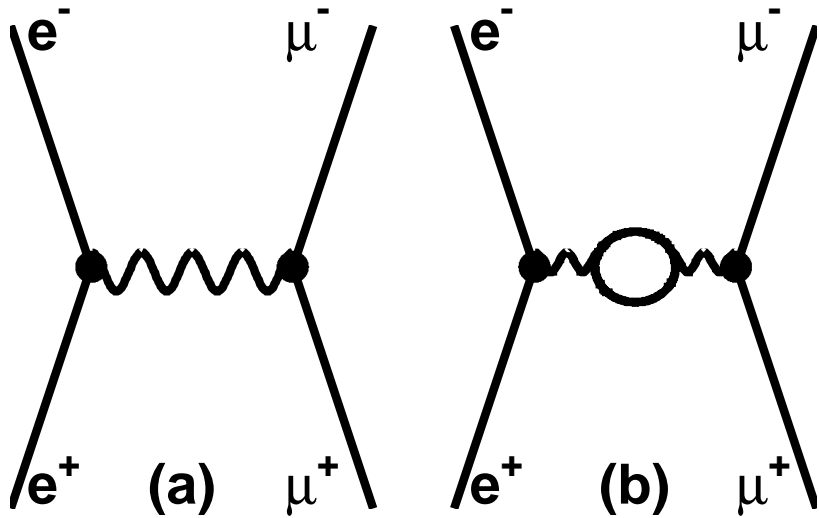,width=15.0cm}
\caption{Feynman diagrams of the process $e^+e^-\to\mu^+\mu^-$. (a) -- diagram 
         in the lowest order, (b) -- vacuum polarization diagram, the loops are
	 due to fermion pairs -- electrons, muons, $\tau$-leptons and quarks.}
\label{mumu-diag}
\end{figure}

 The lowest order Feynman diagram of the process $e^+e^-\to\mu^+\mu^-$ in 
 the energy region $\sqrt{s}<$ 2000 MeV is shown in the Fig.\ref{mumu-diag}(a).
 In Fig.\ref{mumu-diag}(b) the diagram of vacuum polarization containing 
 virtual lepton and quark pairs is also shown. These virtual pairs effectively
 shield a full charge that leads to energy dependence of the electromagnetic 
 coupling constant:
\begin{eqnarray}
 \alpha(s) = {\alpha(0) \over 1-\Pi(s)},
\end{eqnarray}
 where $\Pi(s)$ is the vacuum polarization. The vacuum polarization with 
 leptonic pairs is computed theoretically in the QED framework, while the 
 hadronic vacuum polarization is computed by using dispersion integral and the 
 experimental $e^+e^-\to hadrons$ cross section.

 The process $e^+e^-\to\mu^+\mu^-$ in the energy region $\sqrt{s}<$ 2000 MeV
 was studied earlier in several experiments. In Ref.\cite{borgia,balakin,alles}
 the tests of QED with low statistics were reported. In Ref.\cite{kmd2} the 
 cross section of the $e^+e^-\to\mu^+\mu^-$ process was measured with accuracy
 of about 1\% in the energy region $\sqrt{s}=370$--520 MeV. The studies of the
 $\phi\to\mu^+\mu^-$ decay were reported in Ref.\cite{olya,kloe}.
 
 For the studies of the process $e^+e^-\to\mu^+\mu^-$ with SND detector the 
 most convenient energy region is $\sqrt{s}>980$ MeV. Here the value of the
 $e^+e^-\to\mu^+\mu^-$ process cross section is equal or higher than the
 cross section of the main background process $e^+e^-\to\pi^+\pi^-$ and the
 muons are detected with SND muon system. The SND results of the  
 $\phi\to\mu^+\mu^-$ decay study were published in Ref.\cite{snd1,snd2}. In 
 this work the results of the  $e^+e^-\to\mu^+\mu^-$ process analysis in the
 energy region $\sqrt{s}=980$, 1040 -- 1380 MeV, based on the integrated 
 luminosity $6.4~\mbox{pb}^{-1}$ is presented.

\section{Experiment}

 The SND detector \cite{sndnim} operated from 1995 to 2000 at the
 VEPP-2M \cite{vepp2} collider in the energy range $\sqrt[]{s}$ from 360 to
 1400 MeV. The detector contains several subsystems. The tracking system
 includes two cylindrical drift chambers. The three-layer spherical
 electromagnetic calorimeter is based on NaI(Tl) crystals.
 The muon/veto system consists of plastic scintillation counters and two
 layers of streamer tubes. The calorimeter energy and angular resolutions
 depend on the photon energy as
 $\sigma_E/E(\%) = {4.2\% / \sqrt[4]{E(\mbox{GeV})}}$ and
 $\sigma_{\phi,\theta} = {0.82^\circ / \sqrt[]{E(\mathrm{GeV})}} \oplus
 0.63^\circ$. The tracking system angular resolutions are about
 $0.5^\circ$ and
 $2^\circ$ for azimuthal and polar angles respectively.
	    
\section{Data Analysis}

 The cross section of the $e^+e^-\to\mu^+\mu^-$ process was measured in the
 following way.
\begin{enumerate}
\item
 The collinear events
 $e^+e^-\to\mu^+\mu^-$, $e^+e^-\to e^+e^-$ and $e^+e^-\to\gamma\gamma$
 were selected.
\item
 The $e^+e^-\to e^+e^-$ and $e^+e^-\to\gamma\gamma$ events were used for
 integrated luminosity determination:
\begin{eqnarray}
 IL={{N}\over{\sigma(s)\varepsilon(s)}},
\end{eqnarray}
 where $N$, $\sigma(s)$ and $\varepsilon(s)$  are event number, cross section
 and detection efficiency for the process $e^+e^-\to e^+e^-$ or
 $e^+e^-\to\gamma\gamma$.
\item
 The cross section of the process  $e^+e^-\to\mu^+\mu^-$ was obtained as:
\begin{eqnarray}
 \sigma_{\mu\mu}(s) = {{N}\over{IL\varepsilon(s)\delta_{rad}(s)}}.
\end{eqnarray}
 Here $N$ is the selected events number of the process $e^+e^-\to\mu^+\mu^-$, $IL$
 is integrated luminosity, $\varepsilon(s)$ is the detection efficiency,
 $\delta_{rad}(s)$ is the radiative correction which takes into account the 
 emission of photons by the initial and final particles 
 \cite{KuraevFadin,fedot}.
\end{enumerate}
\begin{figure}[p]
\begin{center}
\epsfig{figure=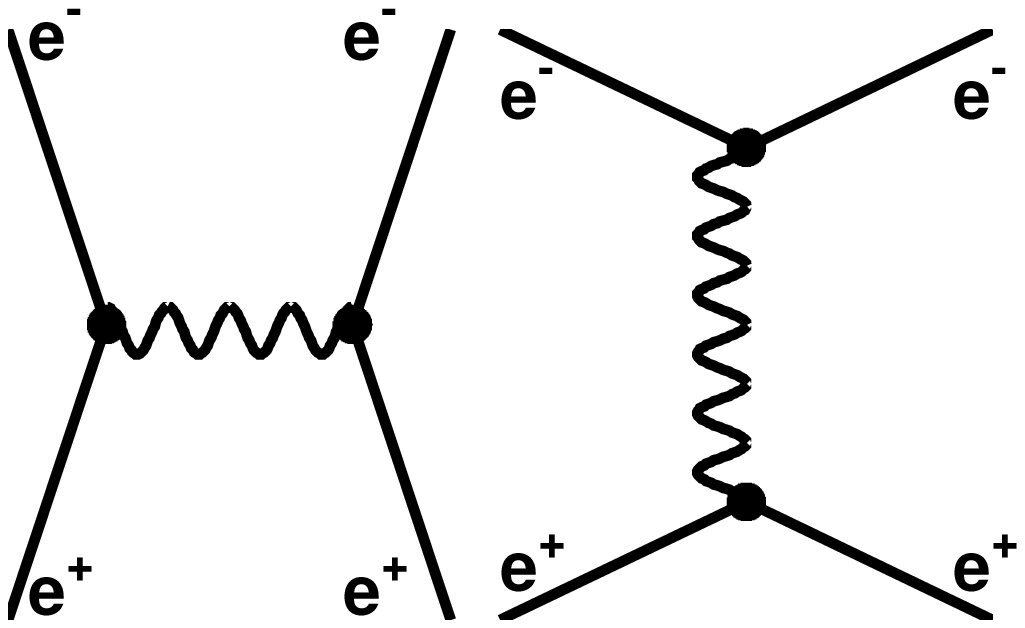,width=15.0cm}
\caption{The Feynman diagrams of the $e^+e^-\to e^+e^-$ process in lowest 
         order.}
\label{ee-diag}
\epsfig{figure=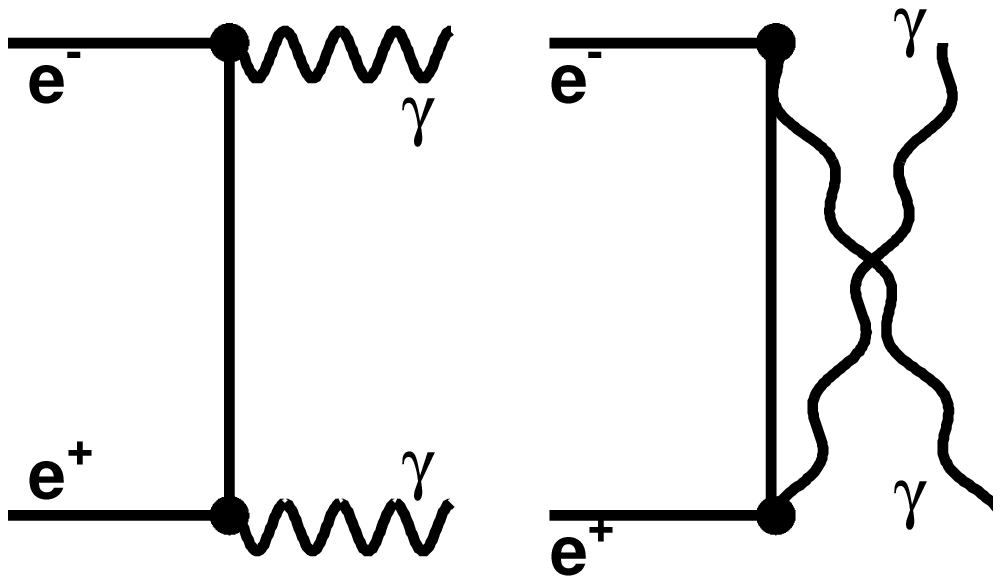,width=15.0cm}
\caption{The Feynman diagram of the $e^+e^-\to\gamma\gamma$ process in lowest 
         order.}
\label{gg-diag}
\end{center}
\end{figure}

\begin{figure}[t]
\begin{center}
\epsfig{figure=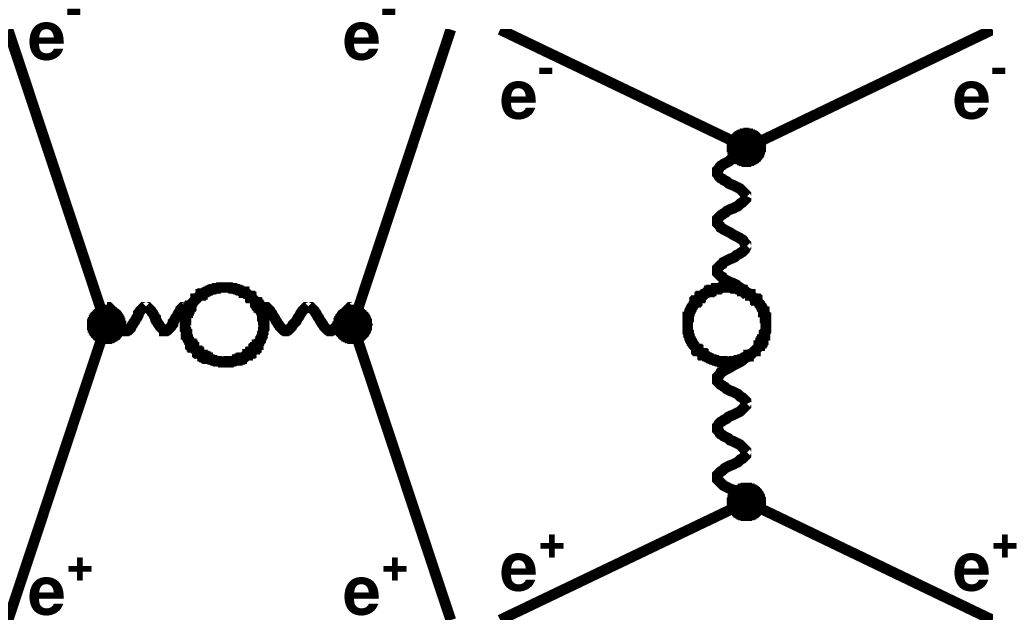,width=15.0cm}
\caption{The Feynman diagrams of the process $e^+e^-\to e^+e^-$ with vacuum
         polarization due to virtual  fermion pairs (electrons, muons, 
	 $\tau$-leptons and  quarks).}
\label{ee-diag-pol}
\end{center}
\end{figure}
 
 The detection efficiency was obtained from Monte Carlo (MC) simulation
 \cite{sndnim,sndpi2}. In order to obtain the detection efficiency of the 
 $e^+e^-\to\mu^+\mu^-$ process, the MC events generator based on the formula
 obtained in the Ref.\cite{arbuzqed} was used. MC simulation of the processes
 $e^+e^-\to e^+e^-$ and $e^+e^-\to\gamma\gamma$ was based on the formulas
 obtained in the Ref.\cite{berklee,berklgg}. The simulation of the process
 $e^+e^-\to e^+e^-$ was performed with the cut 
 $30^\circ<\theta_{e^\pm}<150^\circ$ on the polar angles of the final electron
 and positron. The cross section under these conditions was computed by using
 BHWIDE \cite{bhwide} code with accuracy  0.5 \%.
  
 The Feynman diagrams of the processes $e^+e^-\to e^+e^-$ and 
 $e^+e^-\to\gamma\gamma$ in the lowest order are shown in Fig.\ref{ee-diag} 
 and \ref{gg-diag}. The process $e^+e^-\to e^+e^-$ also contains the
 contribution from the vacuum polarization due to leptons and hadrons virtual 
 pairs (Fig.\ref{ee-diag-pol}), while the process 
 $e^+e^-\to\gamma\gamma$ does not have such contributions. Hence to obtain the
 deviation of $\alpha(s)$ from $\alpha(0)$, the process $e^+e^-\to\gamma\gamma$
 is preferable for normalization.
 
 In this work the cross section of the process $e^+e^-\to\mu^+\mu^-$ was 
 obtained based on integrated luminosities measured by using both 
 $e^+e^-\to e^+e^-$ ($IL_{ee}$) and $e^+e^-\to\gamma\gamma$ 
 ($IL_{\gamma\gamma}$) processes. The cross section of the process 
 $e^+e^-\to e^+e^-$ in the angular region $30^\circ<\theta_{e^\pm}<150^\circ$ 
 was measured by using integrated luminosity
 $IL_{\gamma\gamma}$:
\begin{eqnarray}
 \sigma_{e^+e^-(\gamma)} =
 { {N_{e^+e^-}} \over{IL_{\gamma\gamma}\varepsilon_{e^+e^-}}},
\end{eqnarray}
 where $N_{e^+e^-}$ and $\varepsilon_{e^+e^-}$ are the event number and
 detection efficiency for the process  $e^+e^-\to e^+e^-$.

\subsection{Selection criteria}

 During the experimental runs, first-level trigger selects events of
 various types: events with charged particles and events containing the
 neutral particles only.  In the  first case, the trigger selected events with
 one or more tracks in the tracking system and with two clusters in the 
 calorimeter with the spatial angle between the clusters more than $100^\circ$.
 The threshold on the energy deposition in cluster was equal to 25 MeV. The 
 threshold on the total energy deposition  in the calorimeter was set equal to
 160 MeV. In the second case, the events without tracks in the tracking system
 and with veto signal of the muon system and with total energy deposition more
 than 250 MeV were selected. During processing of the experimental data, the 
 event reconstruction is performed \cite{sndnim,phi98}. The reconstructed 
 particles were sorted in the decreasing order of their energy deposition in 
 the calorimeter. Further the first two  particles were considered. They were 
 numbered in the following way: in odd events the particle which has the higher
 energy deposition in the calorimeter was named the first one and in the even 
 events the first particle was the particle with 
 lower energy deposition. 
 
 The $e^+e^-\to\gamma\gamma$ process events were selected by using the 
 following selection criteria (below subscripts 1 and 2 denote the first and
 second particles respectively):
\begin{itemize}
\item
 $N_{cha}=0$ and $N_{neu}\ge 2$, where $N_{cha}$, $N_{neu}$ are the numbers of
 charged and neutral particles (photons). Extra photons in the
 $e^+e^-\to\gamma\gamma$ events can appear because of overlap with the beam
 background or due to electromagnetic showers splitting. 
\item
 $55^\circ<\theta_1<125^\circ$, where $\theta$ is the particle polar angle 
\item 
 $|\Delta\theta|=|180^\circ-(\theta_1+\theta_2)|<20^\circ$.
\item
 $|\Delta\phi|=|180^\circ-|\phi_1-\phi_2||<5^\circ$, where $\phi$ is the 
 particle azimuthal angle.
\item
  $E_{1,2}/E_0>0.7$, where $E_i$ is the $i$th photon ($i=1,2$) energy 
  deposition, $E_0$ is the beam energy. 
\end{itemize}
 
 The events of the processes $e^+e^-\to e^+e^-$ and $e^+e^-\to\mu^+\mu^-$ were
 selected in the following way:
\begin{itemize}
\item
 $N_{cha}=2$. The events can contain neutral particles due to  overlap with
 the beam background or due to electromagnetic showers splitting.
\item
 $|z_{1,2}|<10$ cm and  $r_{1,2}<1$ cm, where $z$ is the coordinate of the
 charged particle production point along the beam axis (the longitudinal size
 of the interaction region depends on beam energy and varies from 2 to 3 cm),
 $r$ is the distance between the charged particle track and the beam axis in 
 the $r-\phi$ plane.
\item
 $55^\circ<\theta_{1}<125^\circ$.
\item
  $|\Delta\phi|<10^\circ$ and $|\Delta\theta|<10^\circ$.
\item
 The region of $240^\circ<\phi_{1,2}<300^\circ$ was excluded, because this
 sector of the $\phi$ angle was not covered with the muon system.
\item
 $r_1<0.1$ cm or $r_2<0.1$ cm. This cut strongly suppressed the contribution of
 cosmic muons in the events selected as $e^+e^-\to\mu^+\mu^-$. 
\end{itemize}
 The last two selection criteria were not applied in the measurement of the 
 $e^+e^-\to e^+e^-$ process cross section.

 Finally the $e^+e^-\to e^+e^-$ events were selected by using cuts on the 
 particles energy depositions $E_{1,2}/E_0>0.7$. The selection of the
 $e^+e^-\to\mu^+\mu^-$ events was done by using the following cuts
 $E_{1,2}>50$ MeV and $E_{1,2}/E_0<0.7$. In addition, each particle was
 required to fire the scintillation counters of the muon system.

\subsection{Background determination.}

\begin{figure}[t]
\begin{center}
\epsfig{figure=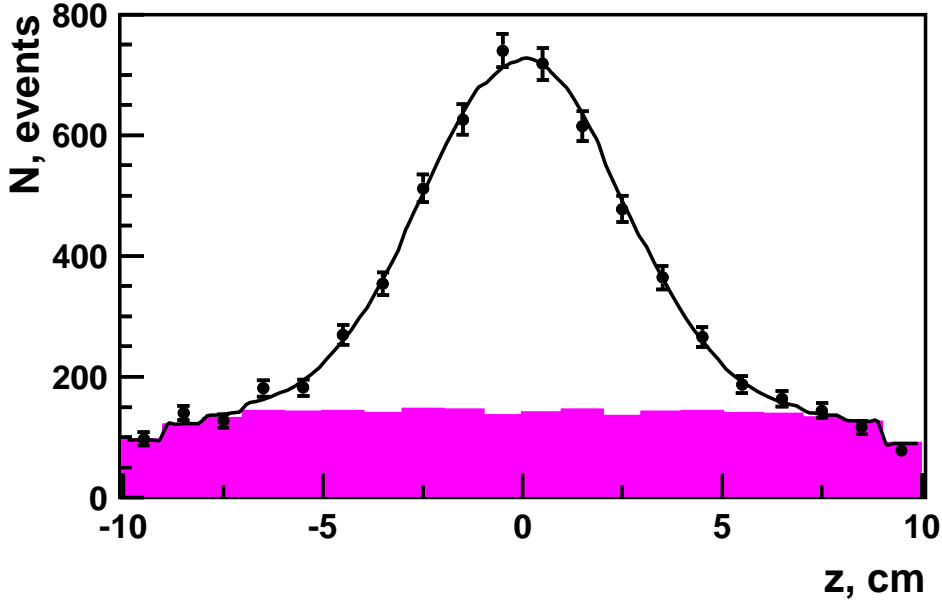,width=15.0cm}
\caption{The distribution of the $z$ coordinate of the charged particles
         production point in events selected as $e^+e^-\to\mu^+\mu^-$ at
	 the energy $\sqrt{s}=1370$ MeV. Dots --  all events, dashed
	 distribution -- cosmic background events, curve -- the fit by sum of
	 distributions of beam and cosmic events.}
\label{koc}
\end{center}
\end{figure}

 The selection criteria described above allow to extract the events of 
 processes $e^+e^-\to e^+e^-$ and $e^+e^-\to\gamma\gamma$ without any
 significant background admixture. The data selected as events of the 
 $e^+e^-\to\mu^+\mu^-$ process contain about  45\% of the cosmic muon 
 background. In order to extract the $e^+e^-\to\mu^+\mu^-$ events number 
 $n_{\mu\mu}$, the distribution over the coordinate  $z=(z_1+z_2)/2$ 
 (Fig.\ref{koc}) was fitted by the sum: 
\begin{equation}
\label{csmfon}
 G(z)\times n_{\mu\mu}+C(z)\times(n-n_{\mu\mu}),
\end{equation}
 where  $n$ is the total number of selected events, $G(z)$ is the Gaussian 
 distribution for $e^+e^-\to\mu^+\mu^-$ events with peak at $z=0$ cm, $C(z)$ 
 is the uniform distribution for cosmic background events.
 The C(z) distribution was obtained by using data collected in special runs
 without beams in collider. The $G(z)$ distribution was obtained in each energy
 point by using $e^+e^-\to e^+e^-$ events. The systematic uncertainty of 
 $n_{\mu\mu}$ determination was estimated by using distributions for the 
 $e^+e^-\to\pi^+\pi^-$ and $K^+K^-$ events instead of $z$-distribution for the
 $e^+e^-\to e^+e^-$ events in eq. (\ref{csmfon}) in the role of $G(z)$. 
 The difference in $n_{\mu\mu}$ values obtained by fitting with various $G(z)$
 was found to be 0.5\% and this value was taken as systematic error due to the
 cosmic background subtraction.
 
 Besides cosmic background, the selected data contain events of the collinear
 $e^+e^-\to e^+e^-$ and $e^+e^-\to\pi^+\pi^-$ processes (the expected 
 background from  the $e^+e^-\to K^+K^-$, $e^+e^-\to 3\pi,4\pi, K_SK_L$
 processes is less than 0.05\%). The expected event number $N_{\pi\pi}$
 from the  $e^+e^-\to\pi^+\pi^-$ process is less than  0.4\% of the 
 $e^+e^-\to\mu^+\mu^-$ event number and was estimated in the following way:
 \begin{eqnarray}
 \label{bg}
 N_{\pi\pi} = \sigma_{\pi\pi}({s}) \varepsilon_{\pi\pi}({s})  IL,
 \end{eqnarray}
 where $\sigma_{\pi\pi}({s})$ is the cross section of the $e^+e^-\to\pi^+\pi^-$
 process measured by OLYA and CMD-2 \cite{olyamh,kmd2mh}, $IL$ is the 
 integrated luminosity , $\varepsilon_{\pi\pi}({s})$ is the detection
 probability for the background process obtained from the simulation under the
 selection criteria described above. The source of error in the $N_{\pi\pi}$
 determination is an inaccurate simulation of the muon system efficiency. To 
 estimate this error the  $e^+e^-\to\pi^+\pi^-$ events were selected at the 
 energy point $\sqrt{s}=980$ MeV (below the $e^+e^-\to K^+K^-$ reaction 
 threshold) by using additional cuts:
\begin{itemize}
\item
 $r_{1,2}<0.1$ cm (for the cosmic background suppression).
\item
 $E_1^{II}<50$ MeV and $E_1^{III}<50$ MeV, where $E_1^{II}$ and $E_1^{III}$
 are the first particle energy depositions in the second and third calorimeter 
 layers respectively (for $e^+e^-\to\mu^+\mu^-$ and cosmic background
 suppression).
\item
 The muon system was not fired by the first particle and no requirements for
 the second particle. 
\end{itemize}
 Under this conditions the $e^+e^-\to e^+e^-$ process background is negligible,
 the cosmic background was subtracted using $z$ distribution. The following 
 value was  obtained:
\begin{eqnarray}
 \delta_{\pi\pi}=\biggl({{n/N}\over{m/M}}\biggr)^2=0.4 \pm 0.4
\end{eqnarray}
 Here $N$ and $M$ are the number of experimental and simulated events of the 
 process $e^+e^-\to\pi^+\pi^-$ selected  under described criteria,
 while $n$ and $m$ are the event numbers in which the muon system was fired by
 the second particle. The accuracy of $\delta_{\pi\pi}$ is equal to its value,
 due to the low statistics.
 
 In the energy region above the $e^+e^-\to K^+K^-$ reaction threshold up to 
 $\sqrt{s}=1100$ MeV the  $\delta_{\pi\pi}$ correction can be obtained by using
 cuts on the $dE/dx$ ionization energy losses in  the drift chamber for the 
 charged kaons background  rejection. In particular at $\sqrt{s}=1100$ MeV it
 was found that $\delta_{\pi\pi}=0.8\pm 0.5$, and this agrees with the 
 estimation presented above.
 
 In order to estimate the systematic uncertainty due to inaccuracy of the
 $N_{\pi\pi}$ subtraction, the probability $\varepsilon_{\pi\pi}({s})$ in all 
 energy points was multiplied by $\delta_{\pi\pi}=0.4$. Then the maximal 
 variation of the measured $e^+e^-\to\mu^+\mu^-$ process cross section was 
 0.7\%.  This value was taken as the systematic error due to the
 $e^+e^-\to\pi^+\pi^-$ background subtraction. 

 The expected value of the  $e^+e^-\to e^+e^-$ events background is about
 0.2\% of the  $e^+e^-\to\mu^+\mu^-$ events number. The systematic error
 due to subtraction of this background was found to be negligible.
 
\subsection{Detection efficiency}
 
 Uncertainties in the simulation of the distributions over some
 selection parameters lead to the inaccuracy in detection efficiency 
 determination. In order to estimate this inaccuracy the experimental and 
 simulated spectra were studied and compared using additional cuts. These cuts
 were selected so that they were uncorrelated with the studied parameter and 
 provided the distribution over this parameters without additional background 
 admixture.
 
 The muon system firing is the main cut for the  extraction of the 
 $e^+e^-\to\mu^+\mu^-$ process events. The comparison of the simulated and experimental probabilities
 of the muon system firing was done by using the following additional cuts:
\begin{itemize}
\item
 $r_{1,2}<0.1$ cm (for the cosmic muon background suppression).
\item
 The muon system was fired by the first particle and no requirements for the
 second particle.
\item
 $30<E_{1,2}^I<55$ MeV, $45<E_{1,2}^{II}<80$ MeV and $55<E_{1,2}^{III}<90$
 MeV, where $E_i^j$ is the ith particle energy deposition in the jth 
 calorimeter layer (for $e^+e^-\to\pi^+\pi^-$ and $K^+K^-$ background 
 rejection).
\end{itemize}
 Then the following parameter was calculated:
\begin{eqnarray}
 \delta_{sc}=\biggl({{n/N}\over{m/M}}\biggr)^2,
\end{eqnarray}
 where $N$, $M$ are selected event numbers and $n$, $m$ are the event numbers
 in which the muon system was fired by the second particle also. The cosmic 
 background was subtracted using $z$-distribution. The coefficient
 $\delta_{sc}$ is equal to 1.15 at $\sqrt{s}=980$ MeV and decreases to 1.0 at
 $\sqrt{s}=1380$ MeV. The detection efficiency of the process
 $e^+e^-\to\mu^+\mu^-$ at various energy points was multiplied by correction 
 coefficient at this point. 
\begin{figure}[t]
\begin{center}
\epsfig{figure=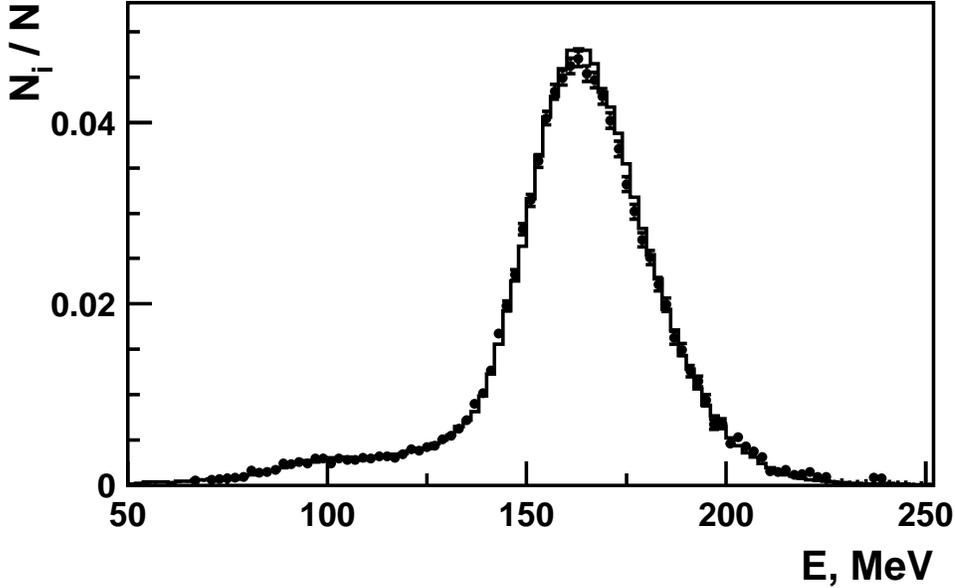,width=15.0cm}
\caption{Energy deposition spectra for the muons in experiment (dots) and
         simulation(histogram).}
\label{mumu-en}
\end{center}
\end{figure}

 The energy deposition spectra of the muons in calorimeter is shown in 
 Fig.\ref{mumu-en}. The experimental and simulated distributions are in good
 agreement. No significant systematics were found due to the cuts on the energy
 deposition in the $e^+e^-\to\mu^+\mu^-$ process.
 
 In the tracking system the particle track can be lost due to reconstruction
 inefficiency. The probabilities to find both tracks were determined by
 using experimental data themselves. It was found to be
 $\varepsilon_{ee}\simeq 0.982\pm 0.001$ and 
 $\varepsilon_{\mu\mu}\simeq 0.983\pm 0.001$ for processes
 $e^+e^-\to e^+e^-$ and $e^+e^-\to\mu^+\mu^-$ respectively. In simulations,
 these values do not actually differ from unity. Thus, if the event numbers 
 of the process $e^+e^-\to\mu^+\mu^-$ were normalized by the integrated 
 luminosity $IL_{ee}$, the systematic errors due to track reconstruction are
 actually reduced. When the events of the  $e^+e^-\to e^+e^-$ and 
 $e^+e^-\to\mu^+\mu^-$ processes were normalized by the luminosity 
 $IL_{\gamma\gamma}$, the detection efficiencies were multiplied by
 coefficients $\varepsilon_{ee}$ and $\varepsilon_{\mu\mu}$.
\begin{figure}[p]
\begin{center}
\epsfig{figure=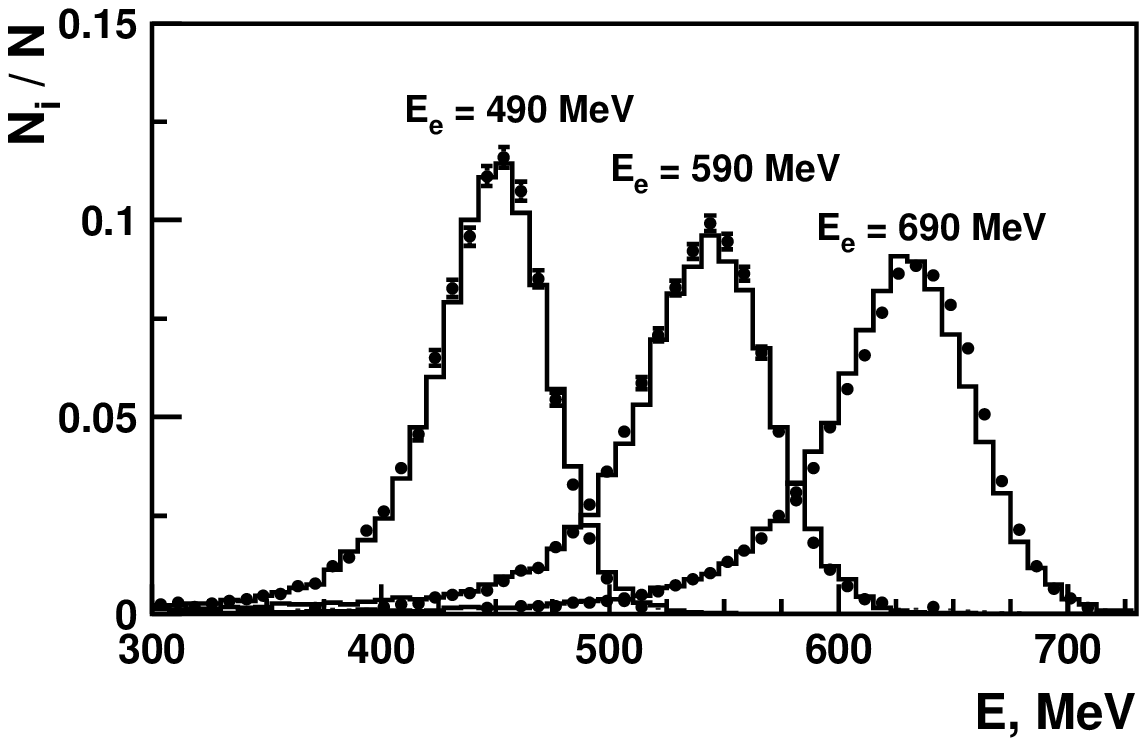,width=15.0cm}
\caption{Energy deposition spectra for electrons with energies of 490, 590 
         and 690 MeV in experiment (dots) and simulation (histogram).}
\label{ee-en}
\epsfig{figure=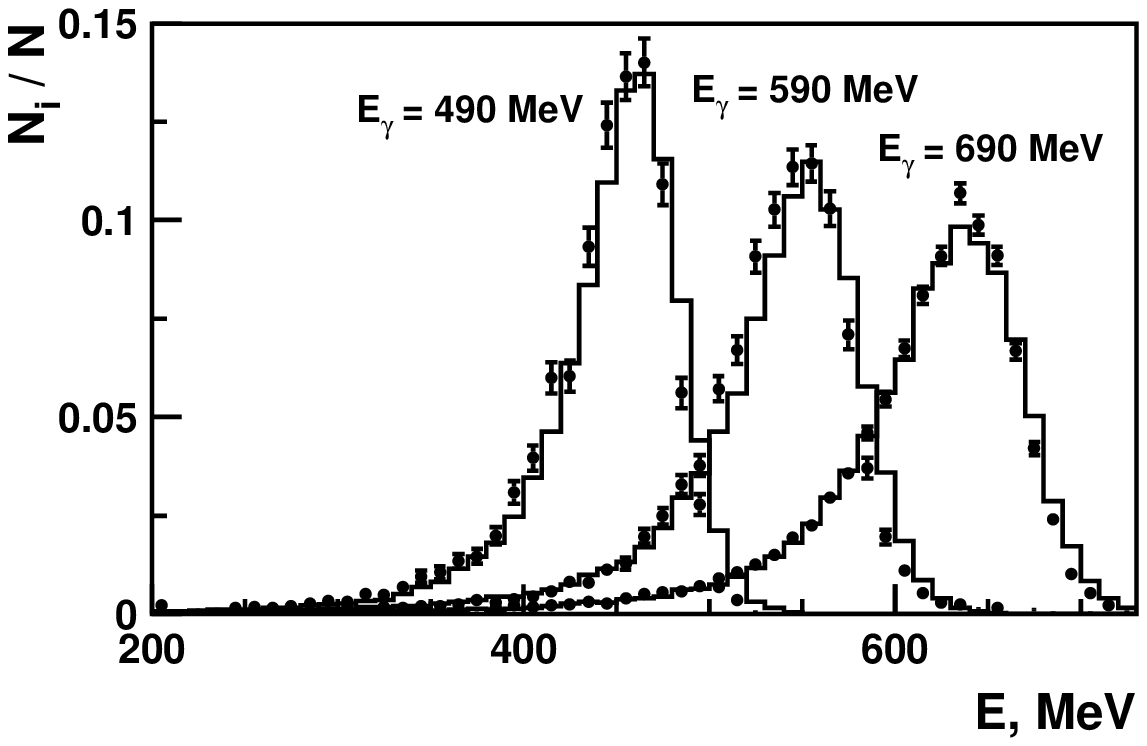,width=15.0cm}
\caption{Energy deposition spectra for photons with energies of 490, 590
         and 690 MeV in experiment (dots) and simulation (histogram).}
\label{gg-en}
\end{center}
\end{figure}
\begin{figure}[p]
\begin{center}
\epsfig{figure=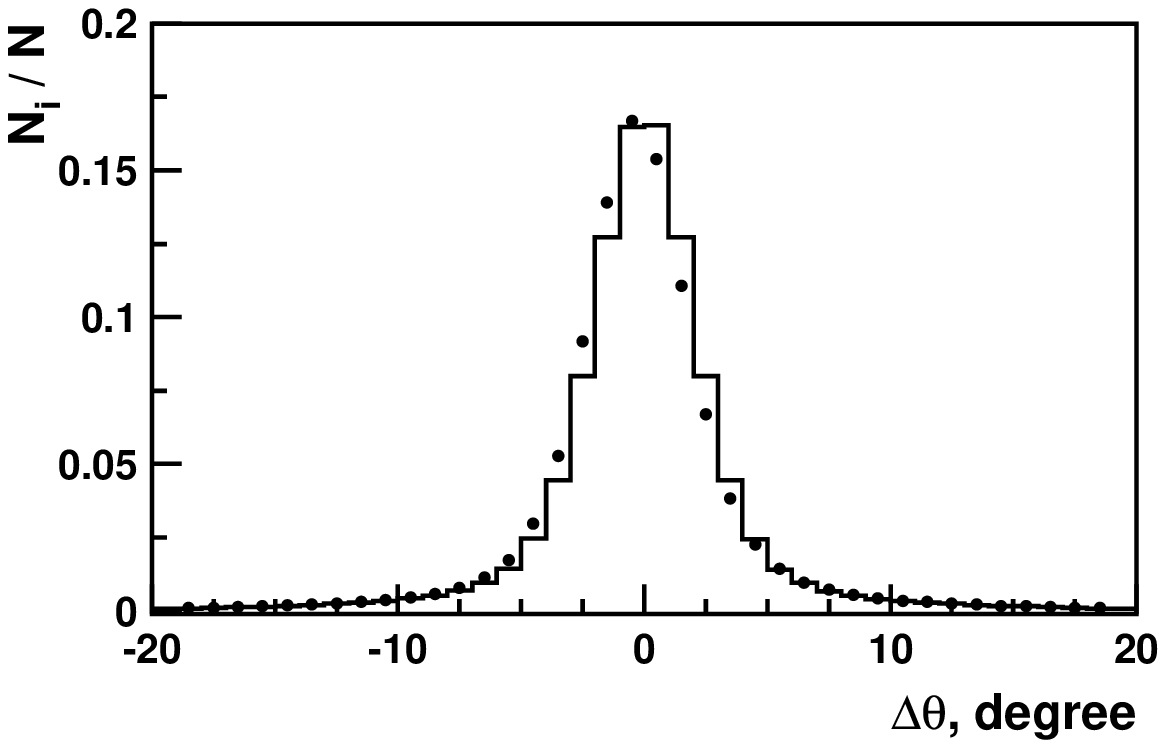,width=15.0cm}
\caption{The $\Delta\theta$ distribution of the $e^+e^-\to e^+e^-$ events. 
         Dots -- experiment, histogram -- simulation.}
\label{dt-ee}
\epsfig{figure=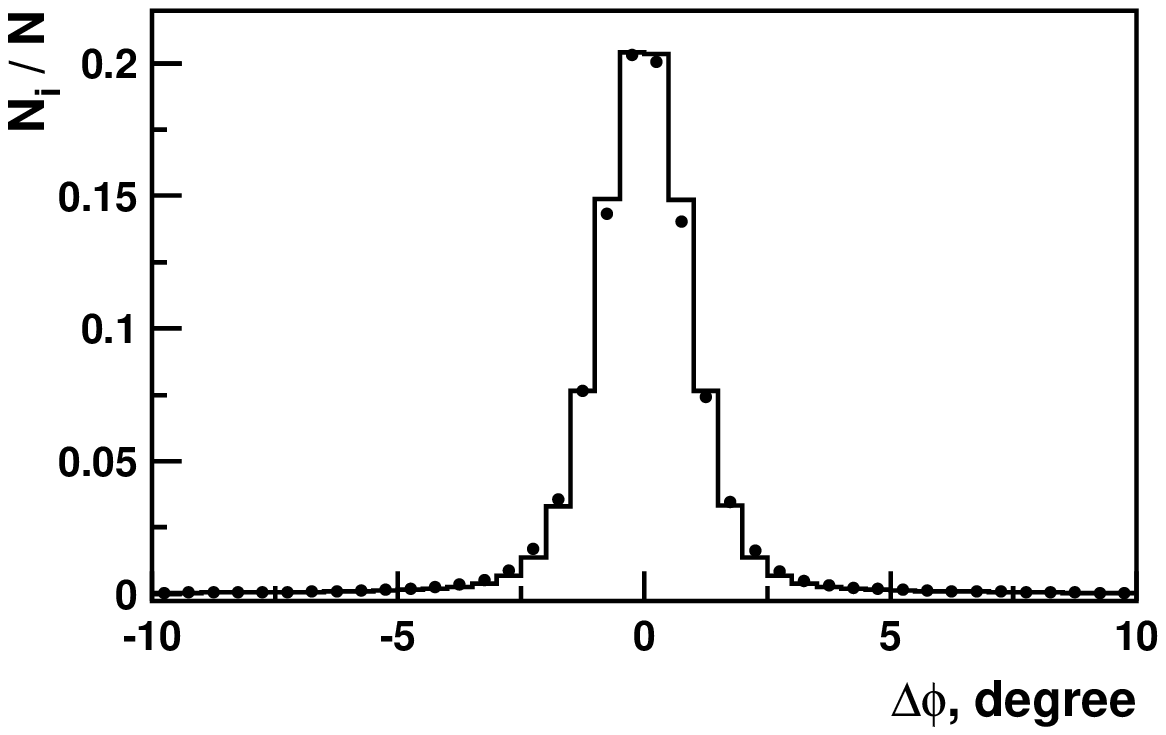,width=15.0cm}
\caption{The $\Delta\phi$ distribution of the $e^+e^-\to e^+e^-$ events. 
         Dots -- experiment, histogram -- simulation.}
\label{df-ee}
\end{center}
\end{figure}
\begin{figure}[p]
\begin{center}
\epsfig{figure=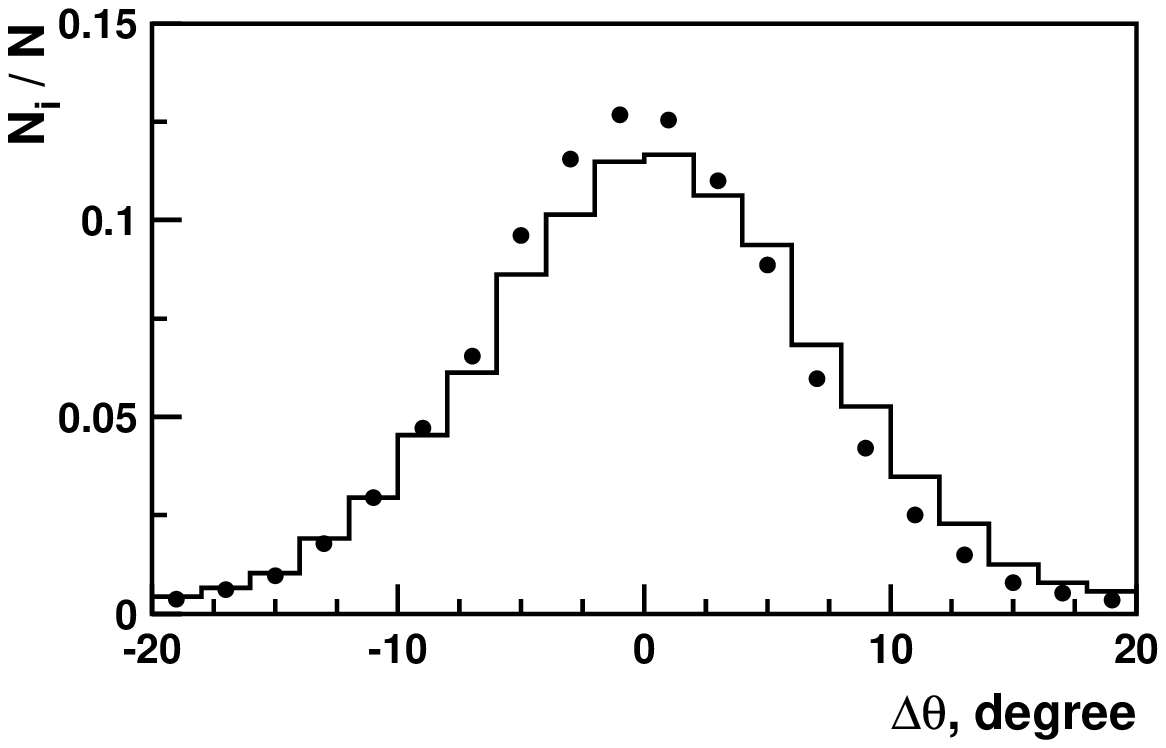,width=15.0cm}
\caption{The $\Delta\theta$ distribution of the $e^+e^-\to\gamma\gamma$ events
         Dots -- experiment, histogram -- simulation.}
\label{dt-gg}
\epsfig{figure=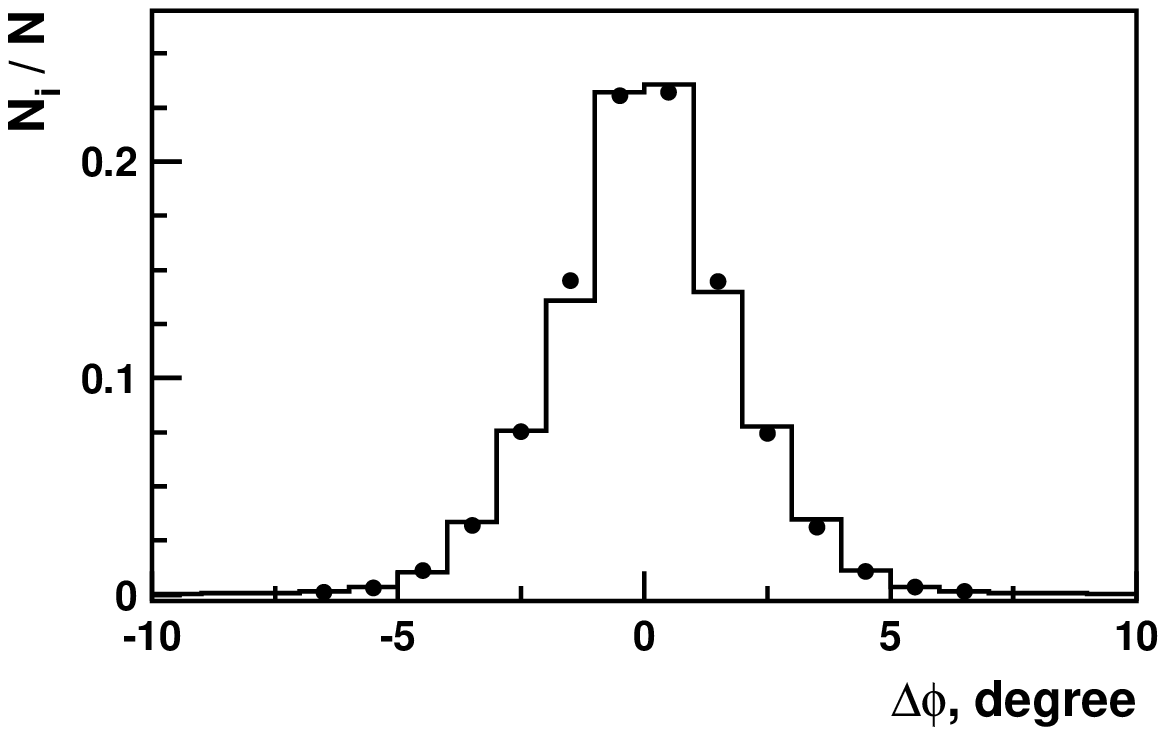,width=15.0cm}
\caption{The $\Delta\phi$ distribution of the $e^+e^-\to\gamma\gamma$ events
         Dots -- experiment, histogram -- simulation.}
\label{df-gg}
\end{center}
\end{figure}
\begin{figure}[p]
\begin{center}
\epsfig{figure=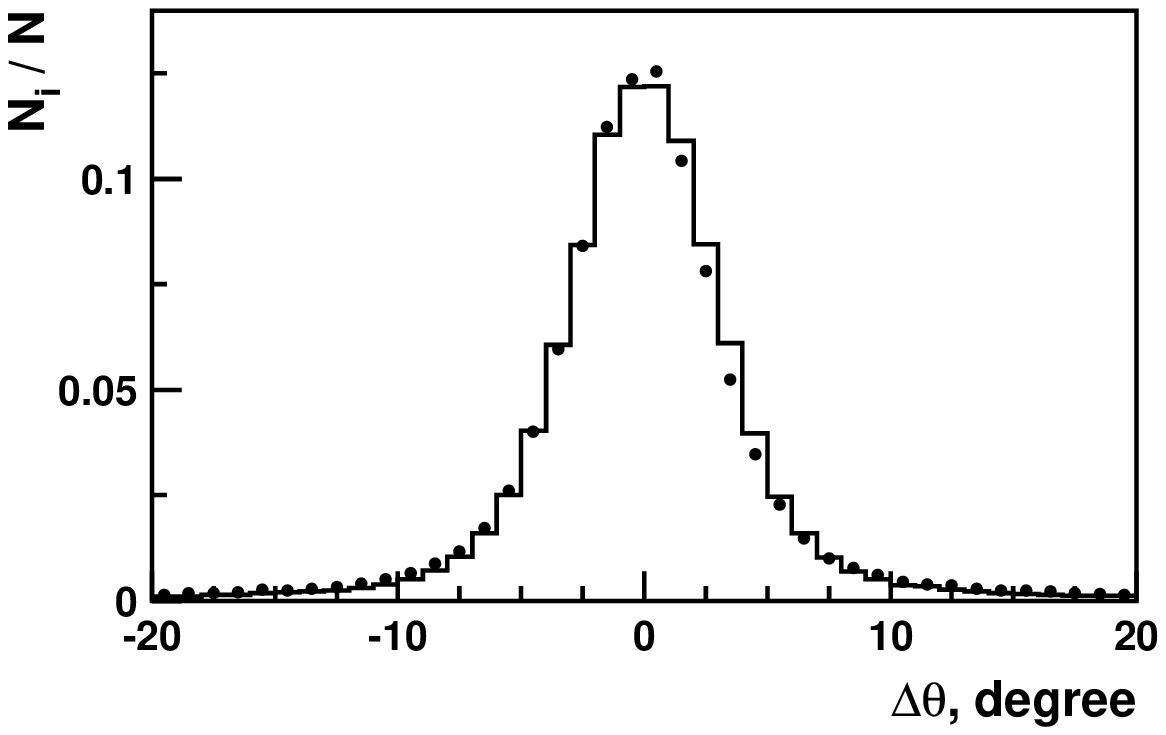,width=15.0cm}
\caption{The $\Delta\theta$ distribution of the  $e^+e^-\to\mu^+\mu^-$ events.
         Dots -- experiment, histogram -- simulation.}
\label{dt-mm}
\epsfig{figure=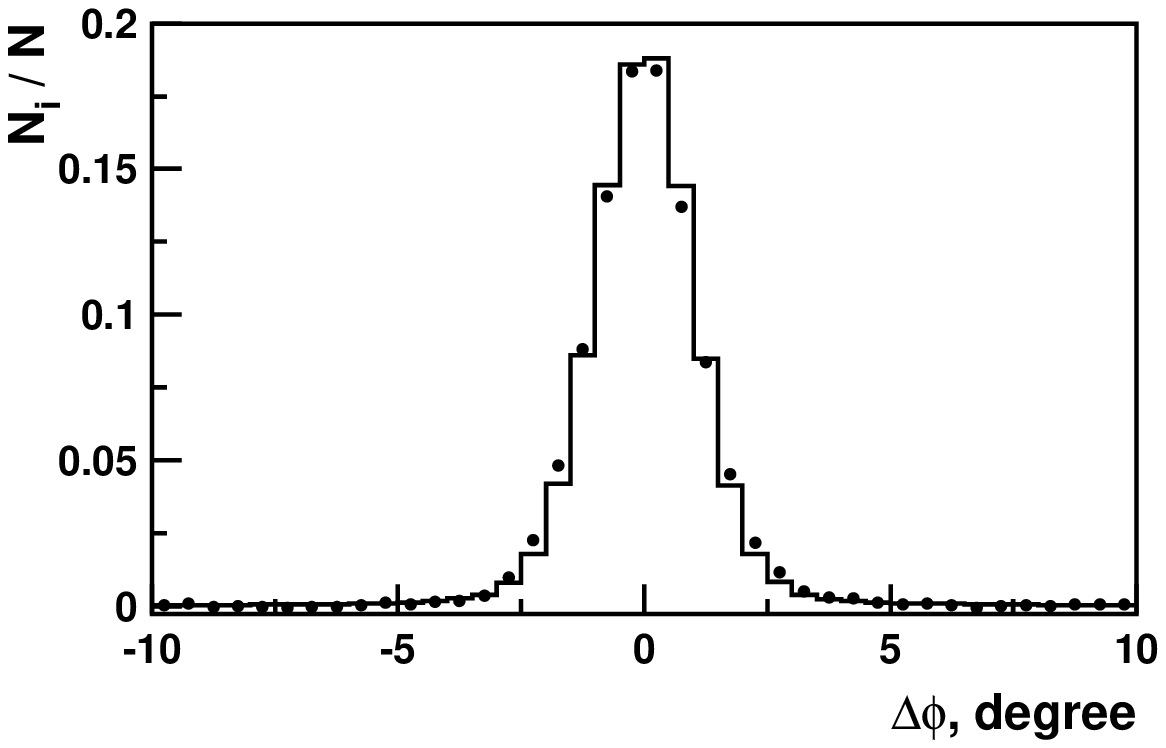,width=15.0cm}
\caption{The $\Delta\phi$ distribution of the $e^+e^-\to\mu^+\mu^-$ events. 
         Dots -- experiment, histogram -- simulation.}
\label{df-mm}
\end{center}
\end{figure}
\begin{figure}[p]
\begin{center}
\epsfig{figure=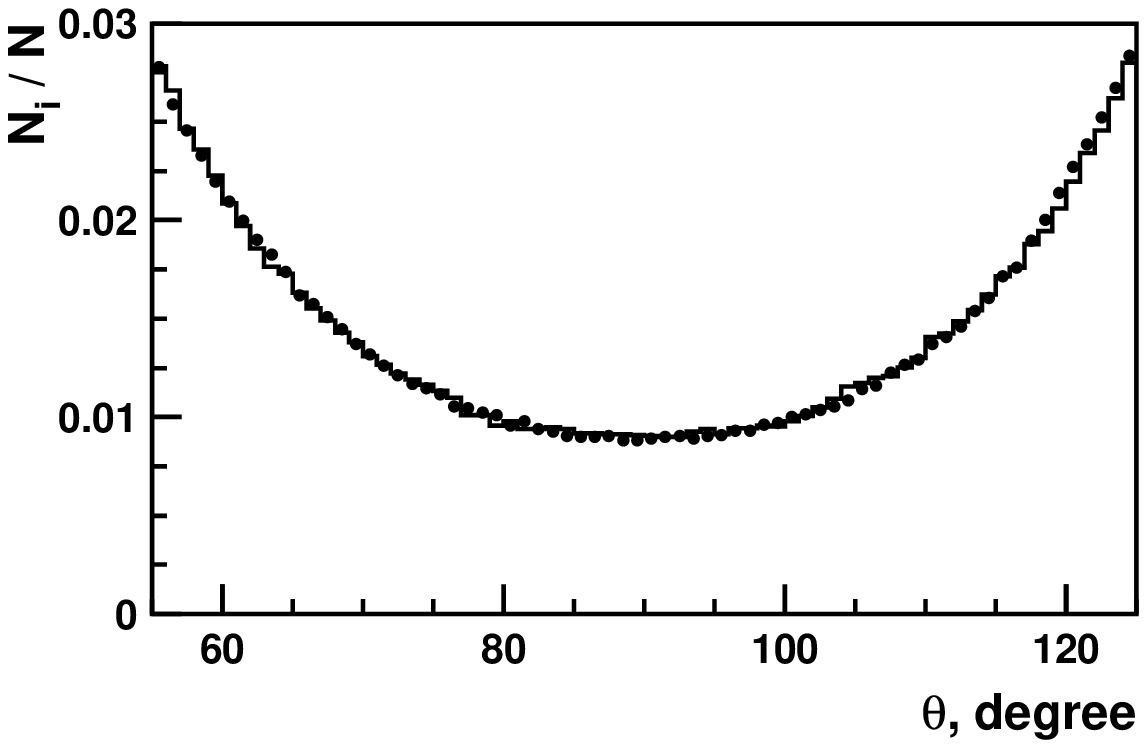,width=15.0cm}
\caption{The $\theta$ angle distribution of the $e^+e^-\to e^+e^-$ events.
         Dots -- experiment, histogram -- simulation.}
\label{tetee}
\epsfig{figure=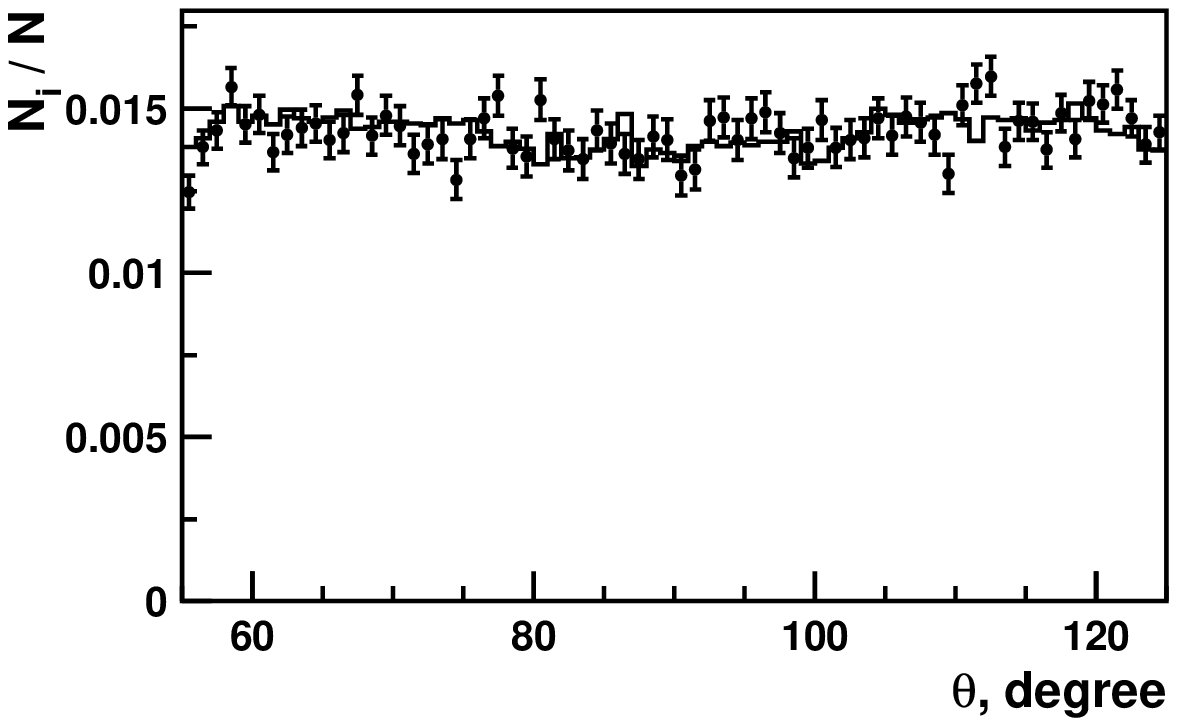,width=15.0cm}
\caption{The $\theta$ angle distribution of the $e^+e^-\to\mu^+\mu^-$ events.
         Dots -- experiment, histogram -- simulation.}
\label{tetmm}
\end{center}
\end{figure}
\begin{figure}[p]
\begin{center}
\epsfig{figure=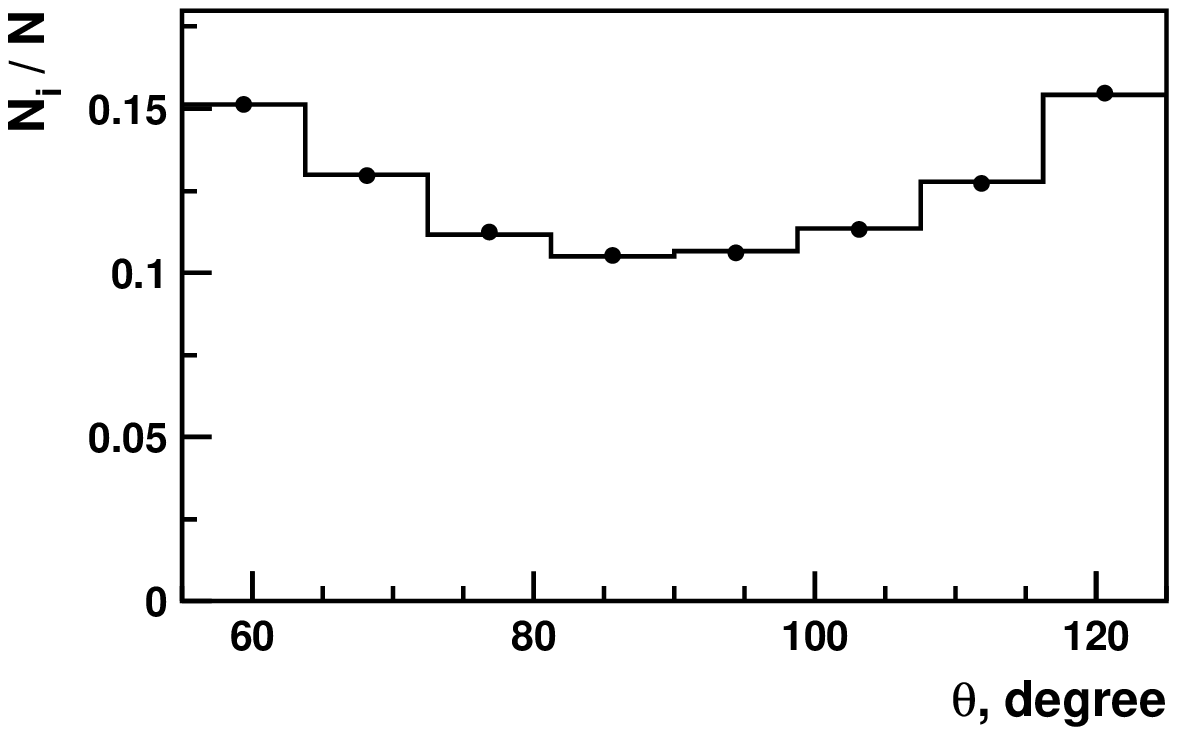,width=15.0cm}
\caption{The $\theta$ angle distribution of the $e^+e^-\to\gamma\gamma$ events.
	 Dots -- experiment, histogram -- simulation.}
\label{tetgg}
\epsfig{figure=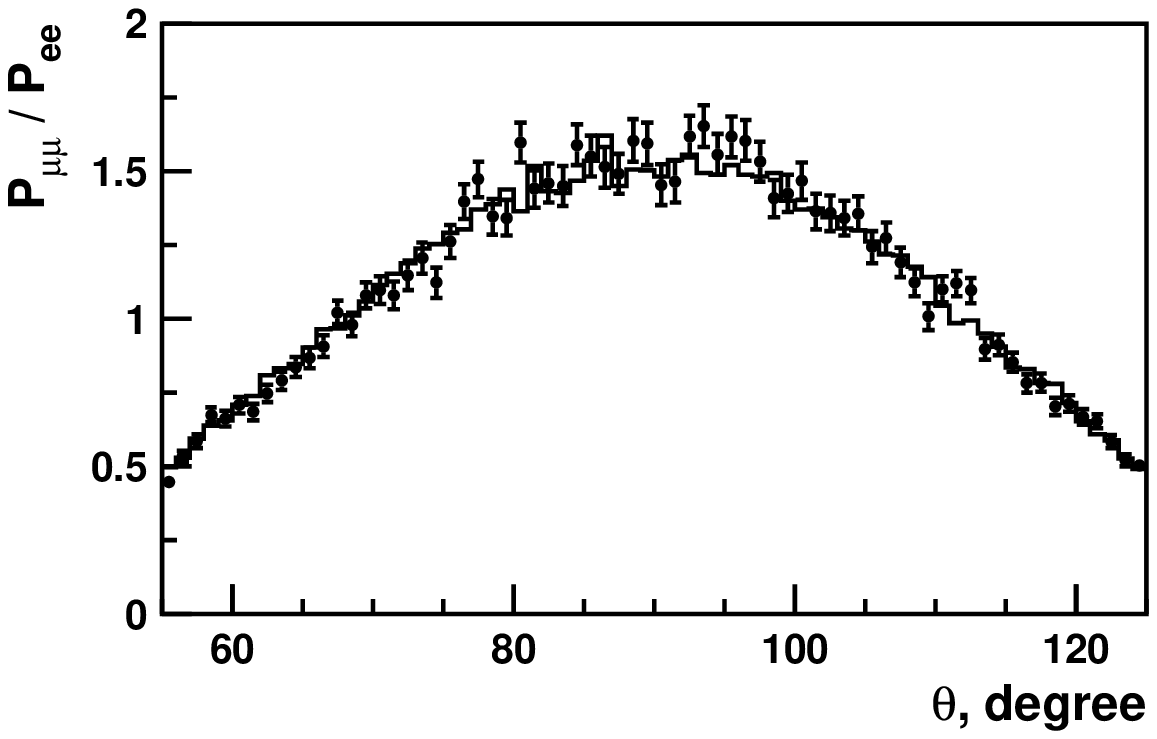,width=15.0cm}
\caption{The ratio of the $\theta$ distributions of the $e^+e^-\to\mu^+\mu^-$ 
         and  $e^+e^-\to e^+e^-$ events. Dots -- experiment, histogram -- 
	 simulation.}
\label{om-mm-ee}
\end{center}
\end{figure}
\begin{figure}[t]
\begin{center}
\epsfig{figure=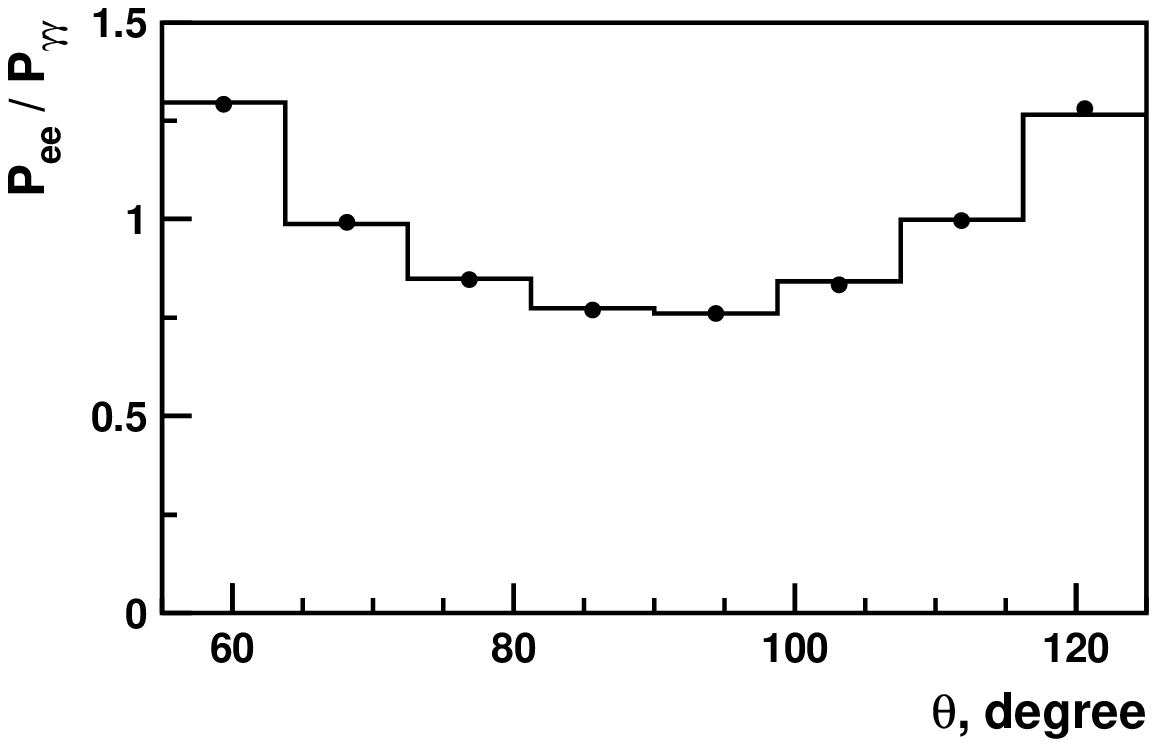,width=15.0cm}
\caption{The ratio of the $\theta$ distributions of the $e^+e^-\to e^+e^-$ and 
          $e^+e^-\to\gamma\gamma$ events. Dots -- experiment, histogram -- 
	  simulation.}
\label{om-ee-gg}
\end{center}
\end{figure}

 The cuts on the $r_{1,2}$ also lead to some inaccuracy of the detection
 efficiency. To obtain the corresponding correction factor to the detection 
 efficiency of the process $e^+e^-\to\mu^+\mu^-$, the events of the process
 $e^+e^-\to\pi^+\pi^-$ were used because in the region $r_{1,2}>0.1$ cm the
 cosmic background dominates and for its rejection  the muon system veto is
 required, which excludes the $e^+e^-\to\mu^+\mu^-$ events also. At the 
 energies under study, muons and pions velocities are about the same and  the 
 drift chamber response on their passage is just the same. In order to exclude 
 the events of the $e^+e^-\to K^+K^-$ process, the correction coefficient was 
 obtained by using data collected at the energy $\sqrt{s}=980$ MeV. MC
 simulation shows that the ratio of event numbers with $r_{1,2}<0.1$ and
 $r_{1,2}>0.1$ is the same for $e^+e^-\to\mu^+\mu^-$ and $e^+e^-\to\pi^+\pi^-$
 processes and does not depend on energy. 
 
 As a result, the detection efficiency of the process $e^+e^-\to\mu^+\mu^-$ was
 multiply by the correction coefficient $\delta^{\mu\mu}_r=0.982\pm 0.005$. The
 error is due to uncertainty of the cosmic muons background subtraction and it
 was added to the systematic error of the detection efficiency. Analogously the
 correction coefficient $\delta^{ee}_r=0.993$ was used for the  $IL_{ee}$ 
 measurement.

 The energy deposition spectra in calorimeter for  $e^\pm$ and $\gamma$ are
 shown in Fig.\ref{ee-en} and \ref{gg-en}. The experimental and simulated
 distributions are in good agreement. The detection efficiency correction
 factor values due to the cuts $E_{1,2}/E_0>0.7$ are usually less than  1\%, but in some 
 energy points it reaches about 3\% and was taken into account for luminosity
 determination. This corrections are the same for both $e^+e^-\to e^+e^-$ and
 $e^+e^-\to\gamma\gamma$ processes -- the average value of the correction
 factors ratio is equal to $1.001\pm 0.001$.

 The $\Delta\phi$ and $\Delta\theta$ distributions of the  $e^+e^-\to e^+e^-$,
 $\gamma\gamma$ and $\mu^+\mu^-$ events are shown in Fig.\ref{dt-ee}, 
 \ref{df-ee}, \ref{dt-gg}, \ref{df-gg}, \ref{dt-mm} and \ref{df-mm}. As a 
 measure of the systematic uncertainty due to the $\Delta\theta$ cut, the 
 following parameter was used:
\begin{eqnarray}
 \delta^{x}_{\Delta\theta}=
 {n_x(|\Delta\theta|<10^\circ)\over N_x(|\Delta\theta|<20^\circ)} \mbox{~} /
 \mbox{~}
 {m_x(|\Delta\theta|<10^\circ)\over M_x(|\Delta\theta|<20^\circ)}, \mbox{~~}
 x=\mu\mu(ee).
\end{eqnarray}
 Here $n_x(|\Delta\theta|<10^\circ)$ and $m_x(|\Delta\theta|<10^\circ)$ are
 the numbers of experimental and simulated events selected under the condition
 $|\Delta\theta|<10^\circ$, while $N_x(|\Delta\theta|<20^\circ)$ and 
 $M_x(|\Delta\theta|<20^\circ)$ are the numbers of experimental and simulated
 events with $|\Delta\theta|<20^\circ$. The average values of
 $\delta_{\Delta\theta}^{ee}$ and $\delta_{\Delta\theta}^{\mu\mu}$ are equal to
 0.999, and have systematic spread of 0.002 and 0.007 respectively. The
 $\delta_{\Delta\theta}^{ee}$ and $\delta_{\Delta\theta}^{\mu\mu}$ were used as
 correction coefficients to the detection efficiencies of corresponding 
 processes.
 
 The variation of the $\Delta\theta$ cut by $5^\circ$ for the 
 $e^+e^-\to\gamma\gamma$ process leads to the variation of the integrated 
 luminosity $IL_{\gamma\gamma}$ by 0.9\%. This value was added to
 the systematic uncertainty of the integrated luminosity measurement.
 Systematic error due to the $\Delta\phi$ cut was found to be negligible for
 all processes.

 The polar angle distributions for the $e^+e^-\to e^+e^-$,
 $e^+e^-\to\mu^+\mu^-$ and $e^+e^-\to\gamma\gamma$ processes are shown in 
 Fig.\ref{tetee}, \ref{tetmm} and \ref{tetgg}. The ratios of these $\theta$
 distributions are shown in Fig.\ref{om-mm-ee} and \ref{om-ee-gg}.
 The experimental and simulated distributions are in good agreement.
 The shapes of the distributions do not depend on energy and for all processes
 is almost the same for the angles $\theta \approx 80^\circ$. Using all
 collected data, the following coefficients were obtained
\begin{eqnarray}
 \delta_\theta^{x}={n_x\over N_x} \mbox{~} / \mbox{~}
 {m_x\over M_x}, \mbox{~~}
 x=\mu\mu(ee,\gamma\gamma),
\end{eqnarray}
 where $N_x$ and $M_x$  are the experimental and simulated event numbers
 in the angular range $55^\circ<\theta<125^\circ$,  while $n_x$ and $m_x$ are
 the experimental and simulated event numbers in the angular range
 $80^\circ<\theta<100^\circ$. In order to estimate the systematic inaccuracy 
 due to the cut on the $\theta$ angle, the following ratio was used:
\begin{eqnarray}
 \delta_\theta={\delta_\theta^{x}\over\delta_\theta^{y}}, \mbox{~~}
   x=\mu\mu(ee), \mbox{~~} y=ee(\gamma\gamma)
\end{eqnarray}
 This ratio was used as the correction factor to the cross section.
 For the process  $e^+e^-\to\mu^+\mu^-$, the $\delta_\theta$ is equal to 
 $1.015\pm 0.010$ and $1.02\pm 0.01$ when it is normalized on the 
 $e^+e^-\to e^+e^-$ and $e^+e^-\to\gamma\gamma$ processes respectively. For 
 the process  $e^+e^-\to e^+e^-$ normalized on the $e^+e^-\to\gamma\gamma$ 
 events, $\delta_\theta=0.995\pm 0.005$. The $\delta_\theta$ error was 
 included in the total systematic error.

 The first-level trigger selection criteria for the $e^+e^-\to\gamma\gamma$ 
 process events included the absence of tracks in the short drift chamber 
 (nearest to the beam-pipe). This leads to the trigger dead time due to the 
 overlap of a background track. The trigger inefficiency of about 5\% was 
 hardware measured during  the data tacking and was taken into account in 
 the analysis.
 
 In $e^+e^-\to\gamma\gamma$ process events, the charged particle can appear
 due to the photon conversion on the detector material before the tracking 
 system. As a measure of the systematic inaccuracy associated to this effect,
 the difference from unity of the following quantity was used:
\begin{eqnarray}
 \delta_{con} = \biggl(1 - {n \over 3N}\biggr) / 
         \biggl(1 - {m \over 3M}\biggr),
\end{eqnarray}
 where $N$ and $M$ are the photon numbers in the experiment and simulation;
 $n$ and $m$ are the photons in the experiment and simulation which had a track
 in the second drift chamber. The  probability to find a track was divided by 
 3 which is the ratio of amounts of matter between the drift chambers and 
 before the tracking system. The result $\delta_{con}=0.998\pm 0.002$ shows
 that the difference between photon conversion probabilities in the experiment
 and simulation does not contribute much in the error of the measurements.

\subsection{Measured cross sections.}

 The cross sections of the process $e^+e^-\to\mu^+\mu^-$ and 
 $e^+e^-\to e^+e^-$ are listed in the Table~\ref{tab1}. 
 The total systematic error of the cross section $\sigma_{\mu\mu}^{ee}$ 
 (obtained by using $IL_{ee}$  luminosity) determination is
 $$
 \sigma_{sys}=\sigma_{eff}\oplus\sigma_{bkg}\oplus\sigma_{rad}
 \oplus\sigma_{IL}=1.6\%.
 $$
 Here $\sigma_{eff}$ is the systematic error of the detection efficiency
 determination, $\sigma_{bkg}$ is the systematic error due to background
 subtraction, $\sigma_{IL}$ is the systematic error of integrated luminosity
 determination due to inaccuracy of the $e^+e^-\to e^+e^-$ cross section
 calculation and $\sigma_{rad}$ is the uncertainty of the radiative
 correction calculation. The magnitudes of various contributions to the
 total systematic error are shown in Table~\ref{tabcuc}. 
\begin{table}
\begin{center}
\caption{The main results of this work. $\sigma_{\mu\mu}^{\gamma\gamma}$ is the
 $e^+e^-\to\mu^+\mu^-$ cross section obtained by using luminosity
 $IL_{\gamma\gamma}$,  $\sigma_{ee(\gamma)}$ is the cross section of the 
 process $e^+e^-\to e^+e^-$  in the angular range 
 $30^\circ<\theta_{e^\pm}<150^\circ$. Only uncorrelated errors are shown. The 
 correlated systematic error $\sigma_{sys}$ is 1.8\% for 
 $\sigma_{\mu\mu}^{\gamma\gamma}$ and 1.1\% for $\sigma_{ee(\gamma)}$.
 \bigskip}
\label{tab1}
\begin{tabular}[t]{ccc} \hline\hline
$\sqrt{s}$ (MeV)&$\sigma_{\mu\mu}^{\gamma\gamma}$, nb &
                 $\sigma_{ee(\gamma)}$, nb \\ \hline
  980&96.3$\pm$5.6&2898$\pm$34\\
 1040&83.6$\pm$4.4&2539$\pm$44\\
 1050&82.3$\pm$4.0&2553$\pm$39\\
 1060&84.1$\pm$3.7&2512$\pm$44\\
 1070&80.6$\pm$3.5&2441$\pm$36\\
 1080&82.7$\pm$3.7&2351$\pm$41\\
 1090&72.8$\pm$3.0&2316$\pm$35\\
 1100&77.8$\pm$3.8&2371$\pm$26\\
 1110&69.7$\pm$2.9&2288$\pm$36\\
 1120&71.6$\pm$3.7&2195$\pm$37\\
 1130&71.6$\pm$2.7&2170$\pm$31\\
 1140&67.6$\pm$3.4&2170$\pm$36\\
 1150&70.7$\pm$4.0&2121$\pm$38\\
 1160&73.0$\pm$3.2&2091$\pm$31\\
 1180&66.8$\pm$3.1&2073$\pm$31\\
 1190&61.8$\pm$2.4&1936$\pm$25\\
 1200&66.8$\pm$4.4&1961$\pm$21\\
 1210&65.6$\pm$2.7&1872$\pm$26\\
 1220&58.5$\pm$2.5&1853$\pm$26\\
 1230&67.6$\pm$3.2&1858$\pm$26\\
 1240&59.3$\pm$2.4&1794$\pm$24\\
 1250&58.8$\pm$2.1&1798$\pm$22\\
 1260&56.7$\pm$2.3&1718$\pm$23\\
 1270&57.9$\pm$1.8&1728$\pm$21\\
 1280&53.5$\pm$1.7&1664$\pm$20\\
 1290&51.1$\pm$1.5&1668$\pm$19\\
\hline
\end{tabular}
\end{center}
\end{table}
\begin{table}
\begin{center}
 Table I: (Continued) \\
\begin{tabular}[t]{ccc} \hline\hline
$\sqrt{s}$ (MeV)&$\sigma_{\mu\mu}^{\gamma\gamma}$, nb &
                 $\sigma_{ee(\gamma)}$, nb \\ \hline
 1300&54.4$\pm$1.6&1638$\pm$18\\
 1310&52.7$\pm$2.0&1608$\pm$21\\
 1320&48.5$\pm$1.9&1556$\pm$19\\
 1330&52.9$\pm$1.8&1538$\pm$18\\
 1340&49.8$\pm$1.7&1586$\pm$20\\
 1350&49.5$\pm$2.1&1507$\pm$18\\
 1360&50.9$\pm$1.7&1541$\pm$19\\
 1370&45.0$\pm$1.9&1476$\pm$18\\
 1380&48.4$\pm$1.5&1459$\pm$15\\
\hline
\end{tabular}
\end{center}
\end{table}
\begin{table}[ccch]
\begin{center}
\caption{Various contributions to the systematic error of the cross sections
 determination. $\sigma_{sys}$ is the total systematic error.}
\label{tabcuc}
\begin{tabular}[t]{lccc} \hline\hline
 the source of the erorr&contribution to 
                    &contribution to 
		    &contribution to \\ 
		    &$\sigma_{\mu\mu}^{ee}$&$\sigma_{\mu\mu}^{\gamma\gamma}$
		    &$\sigma_{ee(\gamma)}$ \\ \hline
 the $\theta$ distribution&1.0 \%&1.0 \%& 0.5 \% \\
 the $r$      distribution&0.5 \%&0.5 \%& --     \\
\hline
 $\sigma_{eff}$   &1.1 \%&1.1 \%& 0.5 \% \\
\hline
 the cosmic background subtraction & 0.5 \% & 0.5 \% & -- \\
 the background from the $e^+e^-\to\pi^+\pi^-$  & & & \\
 process subtraction & 0.7 \% & 0.7 \% & --\\
\hline
 $\sigma_{bkg}$   &0.9 \%&0.9 \% & -- \\
\hline
 $\sigma_{rad}$   &0.5 \%&0.5 \% & -- \\
\hline
 the $\Delta\theta$ distribution in the $e^+e^-\to\gamma\gamma$ process& -- & 
 0.9 \% & 0.9 \% \\
 calculation of the $e^+e^-\to e^+e^-$ &&& \\
 process cross section& 0.5 \%& -- & -- \\
 calculation of the $e^+e^-\to\gamma\gamma$ &&& \\
 process cross section& -- & 0.5 \% & 0.5 \% \\
\hline
 $\sigma_{IL}$    &0.5 \%&1.0 \%& 1.0 \% \\
\hline
 $\sigma_{sys}$   &1.6 \%&1.8 \%& 1.1 \%  \\
\hline
\end{tabular}
\end{center}
\end{table}
 
 The total systematic error of the cross section
 $\sigma_{\mu\mu}^{\gamma\gamma}$ (obtained by using $IL_{\gamma\gamma}$
 luminosity) determination is
$$
 \sigma_{sys}=\sigma_{eff}\oplus\sigma_{bkg}\oplus\sigma_{rad}
 \oplus\sigma_{IL}=1.8\%.
$$
 Here $\sigma_{IL}$ is the systematic error of the integrated luminosity
 determination which includes the inaccuracy of the $e^+e^-\to\gamma\gamma$ 
 cross section calculation and the $\Delta\theta$ angle measurement error. The
 magnitudes of various contributions to the total systematic error are shown
 in Table~\ref{tabcuc}.
 
 The total systematic error of the $\sigma_{ee(\gamma)}$ cross section
 determination (of the $e^+e^-\to e^+e^-$ process in the angular region
 $30^\circ<\theta_{e^\pm}<150^\circ$) is
$$
 \sigma_{sys}=\sigma_{eff}\oplus\sigma_{IL}=1.1\%.
$$
 The magnitudes of various contributions are also listed in Table~\ref{tabcuc}.

\section{Discussion.}

\begin{figure}[p]
\begin{center}
\epsfig{figure=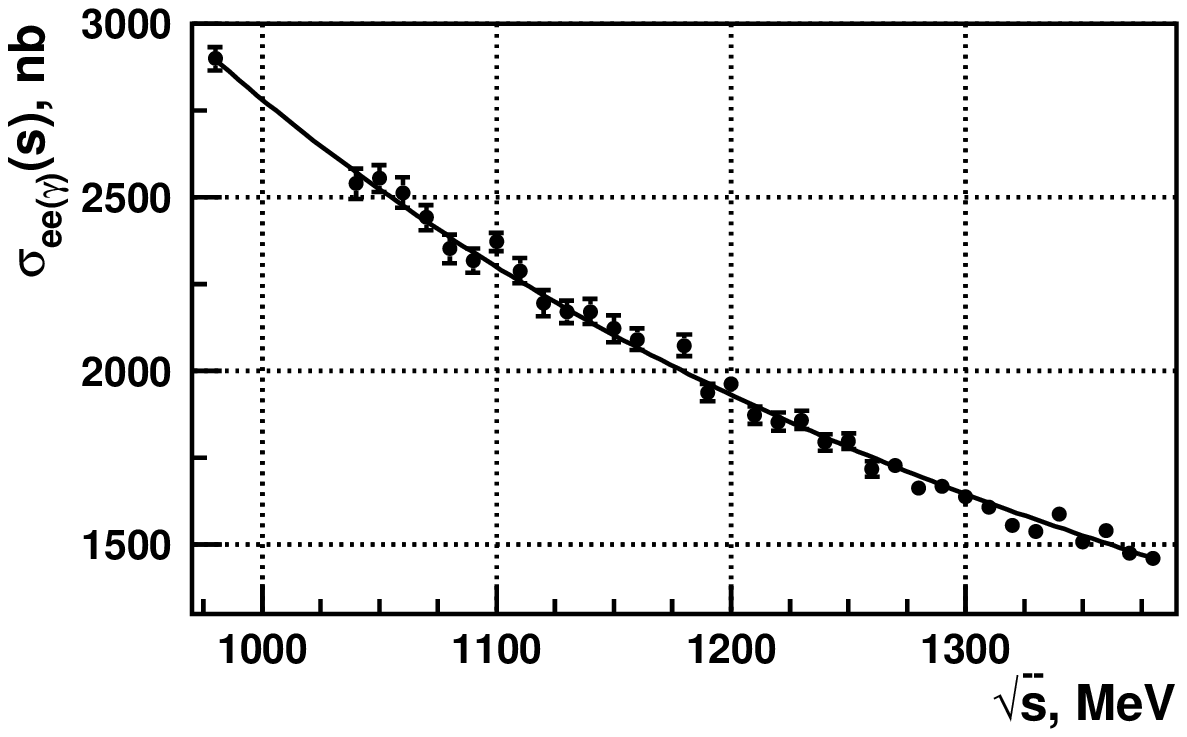,width=15.0cm}
\caption{The $e^+e^-\to e^+e^-$ cross section in the angular range
         $30^\circ<\theta_{e^\pm}<150^\circ$. Dots are the SND data obtained
	 in this work; the curve is the result of the fit
	 ($\chi^2/N_{d.o.f.}=48.1/34$).}
\label{nogro-eeg}
\epsfig{figure=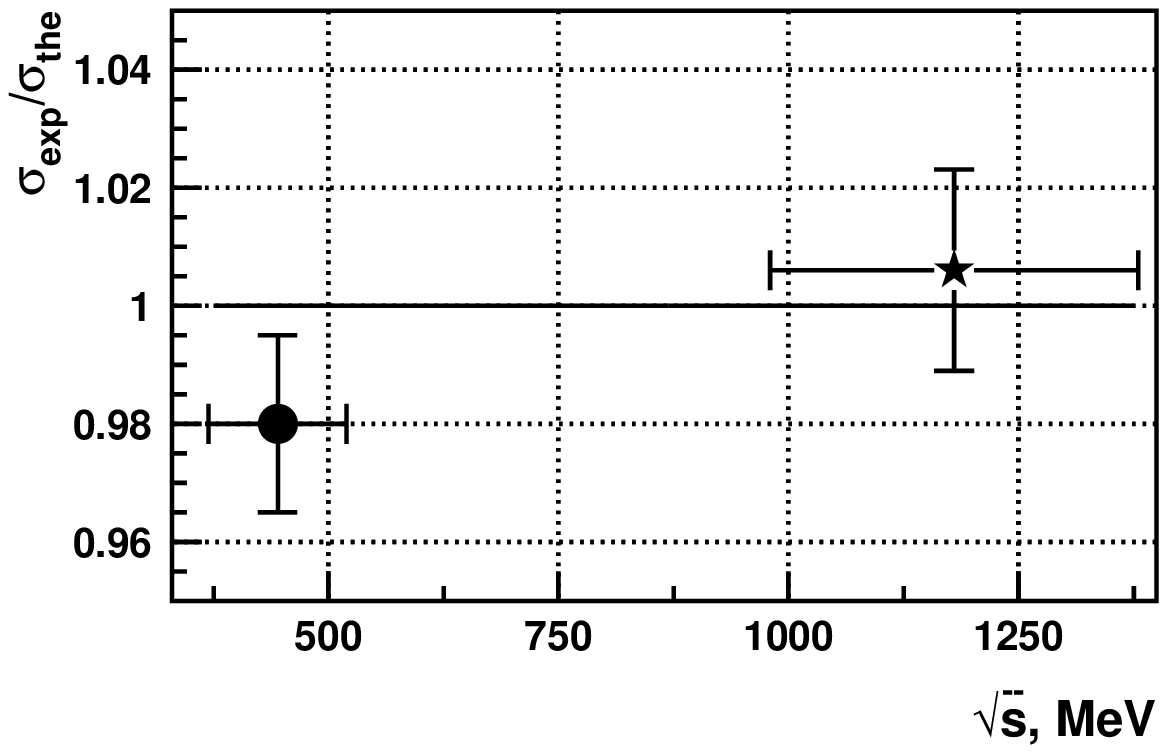,width=13.0cm}
\caption{The ratio $\sigma_{exp}/\sigma_{the}$ of the $e^+e^-\to\mu^+\mu^-$
         cross section measured by SND ($\star$, this work) and CMD-2 
	 ($\bullet$) \cite{kmd2} to theoretical value.
	 The horizontal bars show the energy region $\sqrt{s}$ in which the
         cross section was measured.}
\label{om-snd-kmd}
\end{center}
\end{figure}

 The measured cross section of the process $e^+e^-\to e^+e^-$
 (Table~\ref{tab1}) was fitted with the following expression:
 $$
 \sigma_{ee(\gamma)}=C_{fit} \times {C_{BHWIDE} \over s,}
 $$
 where $C_{BHWIDE}$ is the coefficient calculated by using BHWIDE code 
 \cite{bhwide}. The accuracy of calculation is about 0.5 \%. $C_{fit}$ is the 
 ratio of the measured cross section to theoretically expected (calculated)
 value and it was a free parameter of the fit. As a result, it was obtained
 that (Fig.\ref{nogro-eeg}):
 $$
 C_{fit} = 0.999 \pm 0.002 \pm 0.011
 $$
 The measured value of the $e^+e^-\to e^+e^-$ cross section is in good 
 agreement with calculation.
 
 The $e^+e^-\to\mu^+\mu^-$ cross section was fited with the formula:
\begin{eqnarray}
 \sigma_{\mu\mu} = {4\pi \over 3s} {\alpha(0)^2 \over |1-\Pi(s)|^2}
 {\beta \over 4} \biggl(6-2\beta^2\biggr) \times C_{fit}, \mbox{~~}
 \beta=\sqrt{1-{4m_\mu\over s}}.
\end{eqnarray}
 From the fit of the  $\sigma_{\mu\mu}^{ee}$ cross section, it was found that
 $$
 C_{fit}=1.006 \pm 0.007 \pm 0.016,
 $$
 that agrees well with theoretical predictions.
 
 In the similar energy region  $\sqrt{s}=370$--$520$ MeV, the 
 $e^+e^-\to\mu^+\mu^-$ process cross section was measured by CMD-2 detector
 with accuracy about 1.5 \% \cite{kmd2}. In this experiment the integrated 
 luminosity was obtained by using $e^+e^-\to e^+e^-$ process. The $C_{fit}$ for
 this data was found to be $C_{fit}=0.980\pm 0.013\pm 0.007$. In order to
 compare the SND and CMD-2 results, the following ratio was used:
$$
 C_{fit}^{SND} / C_{fit}^{CMD-2} = 1.027 \pm 0.015 \pm 0.018
$$
 The difference between the SND and CMD-2 results (Fig.\ref{om-snd-kmd}) is
 1.2 standard deviations.
\begin{figure}[p]
\begin{center}
\epsfig{figure=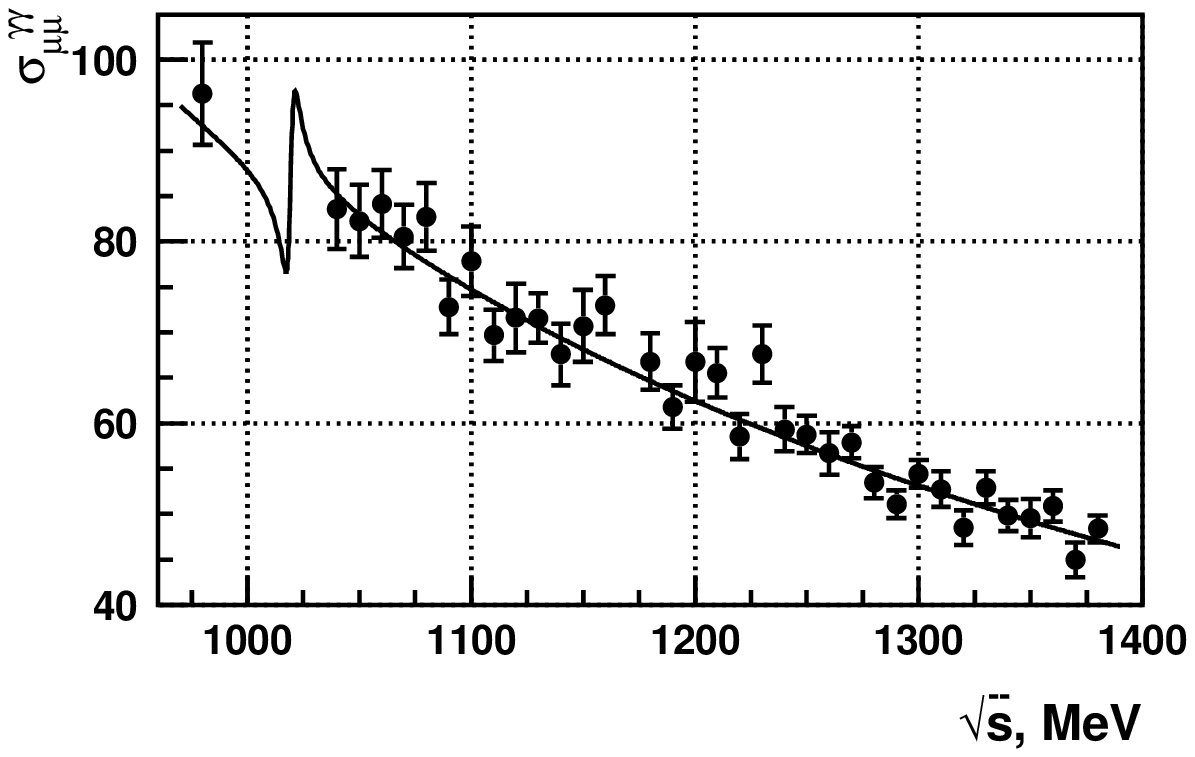,width=15.0cm}
\caption{The $e^+e^-\to\mu^+\mu^-$ cross section obtained by using
         $IL_{\gamma\gamma}$. Dots are the SND data obtained in this work;
	 the curve is the result of the fit ($\chi^2/N_{d.o.f.}=37.2/34$).}
\label{nogro-2}
\epsfig{figure=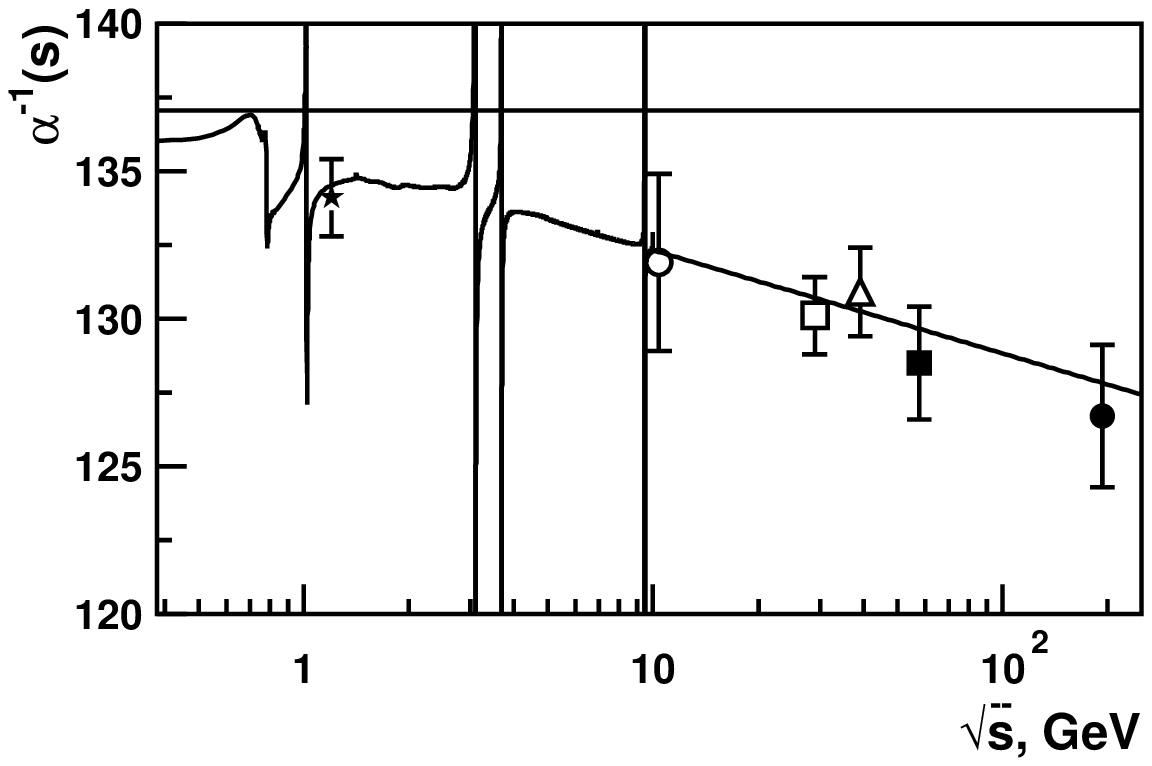,width=12.0cm}
\caption{The $\alpha(s)^{-1}$ values obtained by using different experiments
         results. The SND ($\star$, this work), TOPAZ ($\blacksquare$) 
	 \cite{topaz} and OPAL ($\bullet$) \cite{opal} results are presented.
	 The dots from review \cite{kobel} obtained using results of
	 experiments at DORIS ($\circ$), PEP ($\square$) and PETRA 
	 ($\triangle$) colliders are presented also. Horizontal line shows the
	 $\alpha(0)^{-1}$ value, curve is theoretical calculation of
	 $\alpha(s)^{-1}$.}
\label{alf-2}
\end{center}
\end{figure}
 
 From the fit of the $\sigma_{\mu\mu}^{\gamma\gamma}$ cross section
 (Table~\ref{tab1}, Fig.\ref{nogro-2}) it was found that
$$
 C_{fit}=1.005 \pm 0.007 \pm 0.018.
$$

 If the fit is performed fit with the average value $<1/\alpha>$ as a free 
 parameter, that is by using the function
 to fit data fit function:
\begin{eqnarray}
 \sigma_{\mu\mu} = {4\pi \over 3s} 
 {\beta \over 4} \biggl(6-2\beta^2\biggr) \times 
 \biggl[{1 \over <1/\alpha>} \biggr]^2,
\end{eqnarray}
 then
$$
 <1/\alpha> = 134.1\pm 0.5 \pm 1.2
$$
 This value of $<1/\alpha>$ agrees with expected one, the difference from
 $\alpha(0)$ is  2.3 standard deviations. The obtained value of
 $<1/\alpha>$ together with  other results in the time-like region is shown
 in Fig.\ref{alf-2}. The black markers denote the measurements with
 normalization independent from vacuum polarization diagrams. The results of 
 this work is the only measurement of such a type at the low energy 
 ($\simeq 1$ GeV) region.
 
 A new $e^+e^-$ collider VEPP-2000 for the energy region $\sqrt{s}$ up to
 2 GeV with SND and CMD-2 detectors have now being launching. In the future
 experiments the $e^+e^-\to\mu^+\mu^-$ cross section can be measured with
 accuracy better then  1 \% and it will be a good test of the theory.
 
\section{Conclusion}
 
 The cross section of the process $e^+e^-\to\mu^+\mu^-$ was measured in the
 SND experiment at the VEPP-2M $e^+e^-$ collider in the energy region
 $\sqrt{s}=980,$ 1040 -- 1380 MeV using integrated luminosity obtained from
 the $e^+e^-\to e^+e^-$ and $e^+e^-\to\gamma\gamma$ processes. The accuracy
 of the cross section determination is about 1.6\% and 1.8\% respectively.
 The ratio of the measured cross section to the theoretically predicted value 
 is  $1.006\pm 0.007 \pm 0.016$ and $1.005 \pm 0.007 \pm 0.018$ in the first
 and second case respectively. Using results of the measurements, the
 electromagnetic coupling constant $\alpha$ was obtained in the energy region
 $\sqrt{s}=1040$ -- 1380 MeV: $<1/\alpha> = 134.1\pm 0.5 \pm 1.2$.
 The cross section of the process $e^+e^-\to e^+e^-$ was also measured in the
 angular region $30^\circ<\theta_{e^\pm}<150^\circ$ with systematic accuracy
 1.1\%. The ratio of the measured cross section to
 the theoretically calculated one is $0.999\pm 0.002\pm 0.011$.
  
\acknowledgments
 The work is supported in part by RF Presidential Grant for Sc. Sch.
 NSh-5655.2008.2, RFBR 08-02-00328-a, 08-02-00660-a, 08-02-00634-a, 
 06-02-16192-a, 06-02-16294-a.

\end{document}